\documentclass[sigconf, nonacm]{acmart}

\usepackage{multirow}
\usepackage[linesnumbered,ruled]{algorithm2e}
\usepackage{amsthm}
\usepackage{verbatim}

\theoremstyle{plain}
\newtheorem{theorem}{Theorem}[section]
\usepackage{amsthm}

\newcommand\vldbdoi{XX.XX/XXX.XX}
\newcommand\vldbpages{XXX-XXX}
\newcommand\vldbvolume{14}
\newcommand\vldbissue{1}
\newcommand\vldbyear{2020}
\newcommand\vldbauthors{\authors}
\newcommand\vldbtitle{\shorttitle} 
\newcommand\vldbavailabilityurl{URL_TO_YOUR_ARTIFACTS}
\newcommand\vldbpagestyle{plain} 

\def\eg{\emph{e.g.}} 
\def\ie{\emph{i.e.}}

\begin{document}
\title{Interactive Text-to-SQL via Expected Information Gain for Disambiguation}

\author{Luyu Qiu}
\affiliation{%
  \institution{Hong Kong University of Science and Technology}
  \streetaddress{P.O. Box 1212}
  \city{Hong Kong}
  \state{China}
  \postcode{43017-6221}
}
\email{lqiuag@connect.ust.hk}

\author{Jianing Li}
\affiliation{%
  \institution{Hong Kong Polytechnic University}
  \streetaddress{P.O. Box 1212}
  \city{Hong Kong}
  \state{China}
  \postcode{43017-6221}
}
\email{tensor.li@polyu.edu.hk}

\author{Chi Su}
\affiliation{%
  \institution{Hong Kong Polytechnic University}
  \streetaddress{P.O. Box 1212}
  \city{Hong Kong}
  \state{China}
  \postcode{43017-6221}
}
\email{chisu@polyu.edu.hk}

\author{Lei Chen}
\affiliation{%
  \institution{Hong Kong University of Science and Technology}
  \streetaddress{P.O. Box 1212}
  \city{Hong Kong}
  \state{China}
  \postcode{43017-6221}
}
\email{leichen@cse.ust.hk}

\begin{abstract}
Relational databases are foundational to numerous domains, including business intelligence, scientific research, and enterprise systems. However, accessing and analyzing structured data often requires proficiency in SQL, which is a skill that many end users lack. With the development of Natural Language Processing (NLP) technology, the Text-to-SQL systems attempt to bridge this gap by translating natural language questions into executable SQL queries via an automated algorithm. Yet, when operating on complex real-world databases, the Text-to-SQL systems often suffer from ambiguity due to natural ambiguity in natural language queries. These ambiguities pose a significant challenge for existing Text-to-SQL translation systems, which tend to commit early to a potentially incorrect interpretation.
To address this, we propose an interactive Text-to-SQL framework that models SQL generation as a probabilistic reasoning process over multiple candidate queries. Rather than producing a single deterministic output, our system maintains a distribution over possible SQL outputs and seeks to resolve uncertainty through user interaction. At each interaction step, the system selects a branching decision and formulates a clarification question aimed at disambiguating that aspect of the query. Crucially, we adopt a principled decision criterion based on Expected Information Gain to identify the clarification that will, in expectation, most reduce the uncertainty in the SQL distribution.
This strategy enables globally optimal disambiguation, leading to more accurate and robust query generation with minimal user effort. Moreover, it ensures that each interaction step is interpretable and targeted, enhancing user trust and transparency. Our method aligns with recent advances in interactive and uncertainty-aware semantic parsing and offers a flexible framework adaptable to various real-world datasets and application scenarios. Empirical evaluations demonstrate that our approach significantly improves query accuracy in ambiguous settings while requiring fewer interactions than baselines.
\end{abstract}

\maketitle

\pagestyle{\vldbpagestyle}
\begingroup\small\noindent\raggedright\textbf{PVLDB Reference Format:}\\
\vldbauthors. \vldbtitle. PVLDB, \vldbvolume(\vldbissue): \vldbpages, \vldbyear.\\
\href{https://doi.org/\vldbdoi}{doi:\vldbdoi}
\endgroup
\begingroup
\renewcommand\thefootnote{}\footnote{\noindent
This work is licensed under the Creative Commons BY-NC-ND 4.0 International License. Visit \url{https://creativecommons.org/licenses/by-nc-nd/4.0/} to view a copy of this license. For any use beyond those covered by this license, obtain permission by emailing \href{mailto:info@vldb.org}{info@vldb.org}. Copyright is held by the owner/author(s). Publication rights licensed to the VLDB Endowment. \\
\raggedright Proceedings of the VLDB Endowment, Vol. \vldbvolume, No. \vldbissue\ %
ISSN 2150-8097. \\
\href{https://doi.org/\vldbdoi}{doi:\vldbdoi} \\
}\addtocounter{footnote}{-1}\endgroup

\ifdefempty{\vldbavailabilityurl}{}{
\vspace{.3cm}
\begingroup\small\noindent\raggedright\textbf{PVLDB Artifact Availability:}\\
The source code, data, and/or other artifacts have been made available at \url{\vldbavailabilityurl}.
\endgroup
}

\section{Introduction}

\begin{figure*}
\centering
\includegraphics[width=1\linewidth]{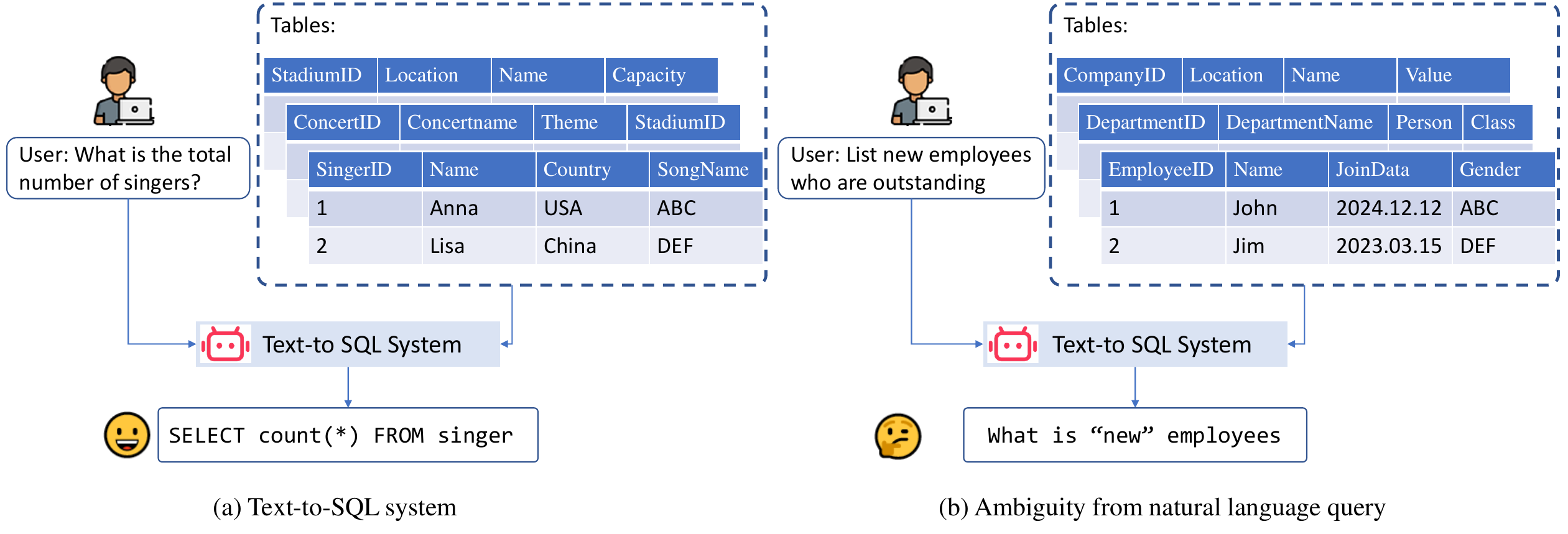}\\
\caption{Illustration of the (a) Text-to-SQL system, and (b) the ambiguity from the natural language query.}
\label{fig:text2sql}
\end{figure*}

\subsection{Natural Language as a Bridge to Databases}

Natural language interfaces serve as a transformative bridge, substantially lowering the barrier for non-expert users who seek to engage with relational databases and perform sophisticated data analysis tasks. The fundamental appeal of these interfaces lies in their intuitive simplicity—users communicate their analytical queries using everyday language, thereby bypassing the need for proficiency in structured query languages such as SQL. Driven by this vision, Text-to-SQL systems~\cite{zhong2017seq2sql,sutskever2014sequence,choi2021ryansql} automate the translation of natural language questions into executable SQL queries, democratizing data access and fostering widespread analytical empowerment. Recent advancements in this area, particularly those leveraging large language models (LLMs), have significantly accelerated progress. Notably, state-of-the-art systems now reach execution accuracy as high as 86.6\% on widely recognized benchmarks like Spider~\cite{gao2023text,yu2018spider}, as illustrated in Fig.~\ref{fig:text2sql}(a).

\subsection{User Intent, Linguistic Expression, and Ambiguity}

Despite impressive empirical performance demonstrated by current Text-to-SQL systems, fundamental challenges persist when deploying these methods in practical, real-world scenarios. At the core, Text-to-SQL translation is not merely a linguistic or syntactic conversion task; it is fundamentally about accurately capturing and translating a user's underlying analytical intent into structured queries. Conceptually, this translation process involves two critical stages: first, users mentally formulate an analytical intention; second, they express this intention using natural language. Most existing Text-to-SQL methodologies predominantly address the second stage—translating the provided natural language into SQL queries—implicitly assuming that the natural language expression completely and unambiguously conveys the user's original intent.

However, this assumption is inherently problematic. Human analytical intentions are often nuanced, context-dependent, and less precise than their linguistic expressions suggest. Natural language, while expressive and flexible, frequently contains ambiguity, underspecification, and implicit contextual dependencies that can obscure the user's true analytical goals. Consequently, even a flawless translation of a natural language query into SQL becomes inadequate if the original linguistic query does not sufficiently encapsulate or clearly represent the user's intended meaning. Addressing the translation process alone, without explicitly recognizing and managing these inherent ambiguities, is insufficient for achieving genuinely accurate and user-aligned query generation in realistic settings.

\subsection{Practical Limitations of Existing Text-to-SQL Systems}

Acknowledging this deeper philosophical limitation highlights two specific and interrelated practical challenges that constrain the effectiveness of current Text-to-SQL systems in realistic deployments. 

Firstly, existing large language models, although powerful, are predominantly trained on generic, web-scale corpora in which SQL knowledge is underrepresented. As evidenced by language usage statistics on platforms such as GitHub, SQL constitutes only a minor fraction of programming data compared to dominant languages like Python. This skewed distribution significantly limits the ability of these models to adequately reason about relational schema, grasp SQL-specific semantics deeply, or robustly generalize to the diverse and complex scenarios commonly encountered in practical database queries.

Secondly, and more fundamentally, is the semantic ambiguity intrinsic to user-generated natural language questions~\cite{bhaskar2023benchmarking,dong2024practiq,wang2023know}. Practical analytical queries posed by users routinely support multiple valid interpretations. For example, a seemingly straightforward request such as \textit{``List new employees"} may implicitly refer to those hired in the last week, the current quarter, or a user-defined interval. Without explicit clarification, purely linguistic models relying solely on textual cues are inherently ill-equipped to disambiguate such uncertainties, frequently producing results misaligned with users' true analytical intentions.

\subsection{Semantic Ambiguity and User Intent}

\begin{figure}
\centering
\includegraphics[width=0.6\linewidth]{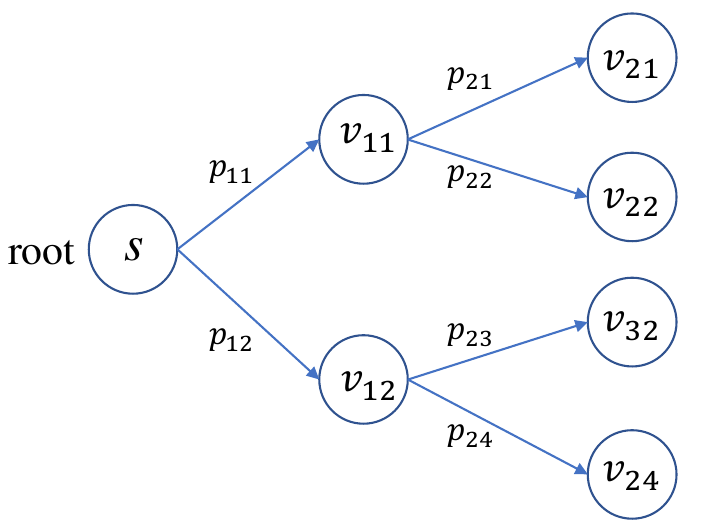}\\
\caption{Illustration of candidate branches.}
\label{fig:branch}
\end{figure}

The ambiguity discussed here is particularly prevalent and problematic in real-world analytical contexts. Unlike schema-related ambiguity arising from database structures, we explicitly address semantic ambiguity embedded within the natural language itself, as depicted in Fig.~\ref{fig:text2sql}(b) and further illustrated by Fig.~\ref{fig:branch}. Such semantic ambiguity manifests through:

\begin{itemize}
    \item Temporal expressions lacking explicit reference points (\textit{e.g.}, ``after 2020," ``recent hires"),
    \item Pronouns and noun phrases with multiple plausible referents (\textit{e.g.}, ``her department"),
    \item Vague qualitative descriptors (\textit{e.g.}, ``top employees" without defined ranking criteria).
\end{itemize}

Consider the illustrative query, \textit{``Show me the top performers from last year."} Here, semantic uncertainty arises from both temporal and evaluative ambiguity: ``last year" may vary in interpretation relative to the current date, fiscal year, or timestamps embedded within the data; simultaneously, ``top performers" could imply multiple evaluation metrics, including sales volume, performance ratings, or customer satisfaction scores.

When faced with these ambiguities, conventional Text-to-SQL systems typically generate a singular, deterministic output~\cite{pilault2023interactive,li2022visa,futeral2022tackling}, often yielding incorrect or unintended results. Such outcomes not only degrade system accuracy but also undermine users' trust, particularly when the generated SQL queries diverge significantly from their actual analytical goals.

\subsection{Interactive Framework: Resolving Ambiguity through User Interaction}

To directly address the inherent semantic ambiguity outlined above, we propose an interactive Text-to-SQL translation framework, designed specifically to treat query interpretation as a collaborative clarification process. Rather than prematurely committing to a single SQL interpretation, our system maintains a probabilistic space of plausible SQL candidates, each corresponding to a different potential resolution of ambiguity inherent within the user's natural language query.

Our approach systematically models SQL generation as an iterative uncertainty-reduction dialogue with the user. At each step, the system identifies specific aspects of the query that exhibit significant semantic variability, such as temporal references, entity interpretations, or constraint criteria, and strategically formulates targeted clarification questions. Crucially, these clarification requests are not arbitrary; instead, the system leverages a principled information-theoretic criterion—specifically, Expected Information Gain (EIG)—to select questions anticipated to most effectively reduce uncertainty regarding user intent.

This targeted interactive approach ensures a globally optimal disambiguation process, enabling the system to converge swiftly and accurately upon the user's intended SQL query. Furthermore, the interactive process maintains transparency, explicitly communicating reasoning to the user, thereby fostering greater trust and interpretability. In stark contrast to conventional one-shot methods reliant on implicit assumptions, our framework aligns explicitly with a user-centric philosophical perspective, emphasizing the necessity of iterative engagement to accurately reflect and refine analytical intentions.

In essence, our proposed interactive Text-to-SQL framework represents both a technical advancement and a philosophical acknowledgment of the inherent complexity and contextuality of human communication, striving toward a deeper alignment between user intent, natural language expression, and structured query generation.

Our contributions can be summarized as follows:

\begin{itemize}
    \item We formally conceptualize Text-to-SQL translation as an interactive, uncertainty-driven inference process, explicitly modeling semantic ambiguity through a probabilistic query interpretation space. This formulation enables the system to iteratively refine and clarify user intent.

    \item We propose a principled interaction strategy based on maximizing the Expected Information Gain (EIG). By leveraging mutual information, our approach strategically selects the most informative clarification questions, leading to effective and interpretable ambiguity resolution.

    \item We implement a comprehensive interactive framework and validate its practical effectiveness through extensive empirical evaluations. Our experiments demonstrate significant improvements in query accuracy and interpretability on benchmark datasets containing realistic semantic ambiguities and complex database schemas.
\end{itemize}

Collectively, our contributions underscore the necessity and value of interactive reasoning within natural language interfaces. Our work not only improves practical usability by aligning query interpretations closely with user intentions but also opens promising avenues for incorporating formal methods of uncertainty quantification into the broader field of semantic parsing.

\section{Formal Definition}

Having established the significance of ambiguity in natural language queries and introduced our interactive Text-to-SQL approach, a critical next step is to precisely formalize this interactive disambiguation process. In this section, we clearly define the interactive Text-to-SQL task, providing rigorous mathematical grounding and systematic explanations that clarify our approach. To effectively handle ambiguity, we introduce a principled framework built upon Expected Information Gain (EIG), enabling our model to intelligently pinpoint which clarifications are most valuable at each interaction. By formalizing this approach, we lay a solid theoretical foundation for achieving efficient, interpretable, and user-aligned Text-to-SQL systems, thereby bridging the gap between theoretical robustness and real-world usability.


The fundamental goal of a Text-to-SQL translation system is to interpret a user's natural language query and convert it into a structured SQL query executable over a relational database. Formally, we define the standard Text-to-SQL task as follows:

\begin{definition}[Text-to-SQL Task]
Given a natural language question \( q \) and a relational database schema \( S = \{T, C, R\} \), consisting of tables \(T\), columns \(C\), and foreign key relations \(R\), the goal of a Text-to-SQL system is to generate a structured SQL query \( y \in \mathcal{Y} \), such that executing \( y \) on the corresponding database instance \(\mathcal{D}_S\) returns the correct answer to the natural language question \( q \). Mathematically, the process can be represented as:
\[
q, S \rightarrow y, \quad \text{where} \quad y(\mathcal{D}_S) = \text{Answer}(q).
\]
\end{definition}

\noindent\textbf{Example 1 (Clear Query).} Consider a straightforward scenario illustrated in Fig.~\ref{fig:text2sql}(a), where a user issues an explicit query:
\[
q = \text{``What is the total number of singers?"}
\]

Given the schema \( S \) that includes a table `singer', a standard Text-to-SQL system accurately translates this query into a deterministic SQL statement:
\[
y = \text{SELECT COUNT(*) FROM singer;}
\]

Modern Text-to-SQL approaches frequently utilize deep neural networks, especially transformer-based architectures or large language models (LLMs), to perform this translation task. Formally, this SQL generation process can be described by the following definition:

\begin{definition}[SQL Generation Model]
Given a natural language question \( q \) and a database schema \( S \), a neural SQL generation model \( F_{\theta} \), parameterized by weights \( \theta \), outputs an executable SQL query \( Q \):
\[
Q = F_{\theta}(q, S).
\]
The model \( F_{\theta} \) typically comprises transformer-based encoder-decoder frameworks, or pretrained large language models (e.g. GPT-4~\cite{achiam2023gpt}, DeepSeek~\cite{liu2024deepseek}).
\end{definition}

However, in real-world applications, natural language queries frequently contain inherent ambiguities arising from linguistic vagueness, underspecification, or context-dependency. Such ambiguities prevent the straightforward mapping of a natural language query to a single definitive SQL statement, thereby requiring explicit modeling. Consider the ambiguous scenario shown in Fig.~\ref{fig:text2sql}(b):

\noindent\textbf{Example 2 (Ambiguous Query).} A user queries:
\[
q = \text{``List new employees who are outstanding."}
\]

The word ``new" introduces ambiguity, as it may refer to:
\begin{itemize}
    \item Employees who joined within a specific recent period (e.g., last month).
    \item Employees hired within the current calendar year.
    \item Employees marked as "new" by an internal company criterion.
\end{itemize}

Thus, we define the notion of ambiguity formally as follows:

\begin{definition}[Ambiguous Query Interpretation]
Given an ambiguous natural language query \( q \) and schema \( S \), the query does not correspond to a single SQL query but instead admits a set of multiple plausible SQL interpretations, denoted as:
\[
\mathcal{Y} = \{(Q_1, p_1), (Q_2, p_2), \dots, (Q_N, p_N)\},
\]
where each candidate SQL query \(Q_i\) represents one interpretation of the query \(q\), and \(p_i = P(Q_i \mid q, S)\) is the probability indicating the likelihood of interpretation \(Q_i\).
\end{definition}

\noindent\textbf{Example 3 (Multiple Interpretations).} Consider the ambiguous natural language query:
\[
q = \text{``List employees who joined after 2020 in sales."}
\]

This simple-sounding query inherently possesses multiple valid SQL interpretations, each corresponding to distinct semantic understandings:

\begin{align*}
Q_1 &: \text{SELECT * FROM employees WHERE join\_date} > \text{'2020-12-31'}\\
&\quad\quad\quad\text{AND department = 'sales'};\\
Q_2 &: \text{SELECT employee\_id, name FROM employees WHERE } \\&\quad\quad\quad \text{join\_date}\geq \text{'2020-01-01'}\text{AND department = 'sales'};\\
Q_3 &: \text{SELECT employee\_id, name FROM employees WHERE } \\
&\quad\quad\quad\text{join\_date}\geq \text{'2021-01-01' AND job\_title LIKE '\%sales\%'};\\
Q_4 &: \text{SELECT * FROM employees WHERE YEAR(join\_date)} = 2020\\
&\quad\quad\quad\text{AND department IN ('sales', 'marketing')};\\
Q_5 &: \text{SELECT employee\_id, name FROM employees WHERE } \\
&\quad\quad\quad\text{join\_date}> \text{'2020-12-31' AND (department = 'sales' OR}\\
&\quad\quad\quad\text{ role = 'sales associate')};\\
&\quad\quad\quad\quad\quad\quad\quad\quad \vdots
\end{align*}

Each candidate query \( Q_i \) above reflects a distinct interpretation arising from linguistic ambiguity:

\begin{itemize}
    \item \textbf{Temporal ambiguity:} "After 2020" could be interpreted strictly after the end of 2020 (\(> \text{'2020-12-31'}\)), or inclusive starting at the beginning of 2020 (\(\geq \text{'2020-01-01'}\)).
    \item \textbf{Scope of "sales":} The query could imply employees working specifically in a department named "sales", those whose job title includes "sales", or roles broadly associated with sales functions.
    \item \textbf{Output schema ambiguity:} "List employees" might refer to retrieving all columns or only a subset of key identifiers such as employee ID and name.
\end{itemize}

Each of these plausible interpretations carries associated probabilities derived from the model's uncertainty, and addressing these explicitly through interactive clarification is necessary for accurate and user-aligned SQL query generation.

Since resolving ambiguity is crucial for practical deployment, our framework introduces an optimal selection criterion based on information theory, specifically the Expected Information Gain (EIG):

\begin{definition}[Optimal Clarification via Information Gain]
Given a set of ambiguous decision variables \(\mathcal{X} = \{X_1, X_2, \dots, X_M\}\), each representing an uncertain interpretation within the query, the optimal decision variable \(X^*\) to clarify first is the one with the highest expected information gain.
where the expected information gain \(I(X_i; \mathcal{Y})\) measures the expected reduction in entropy upon clarifying the decision variable \(X_i\), defined as:

\begin{equation}
I(X_i;\mathcal{Y}) = H(\mathcal{Y}) - H(\mathcal{Y} \mid X_i),
\label{eq:gain}
\end{equation}
with entropy \(H(\mathcal{Y})\) of the query distribution defined as:

\begin{equation}
H(\mathcal{Y}) = -\sum_{y \in \mathcal{Y}} P(y)\log P(y).
\label{eq:entropy}
\end{equation}
\end{definition}

And $H(\mathcal{Y}|X_i)$ is the expected uncertainty entropy after $X_i$ is clarified:
\begin{equation}
    H(\mathcal{Y}|X_i) = \sum_{x\in \mathcal{V}_i}P(x)H(\mathcal{Y}|X_i=x)
\end{equation}

\begin{figure*}
\centering
\includegraphics[width=1\linewidth]{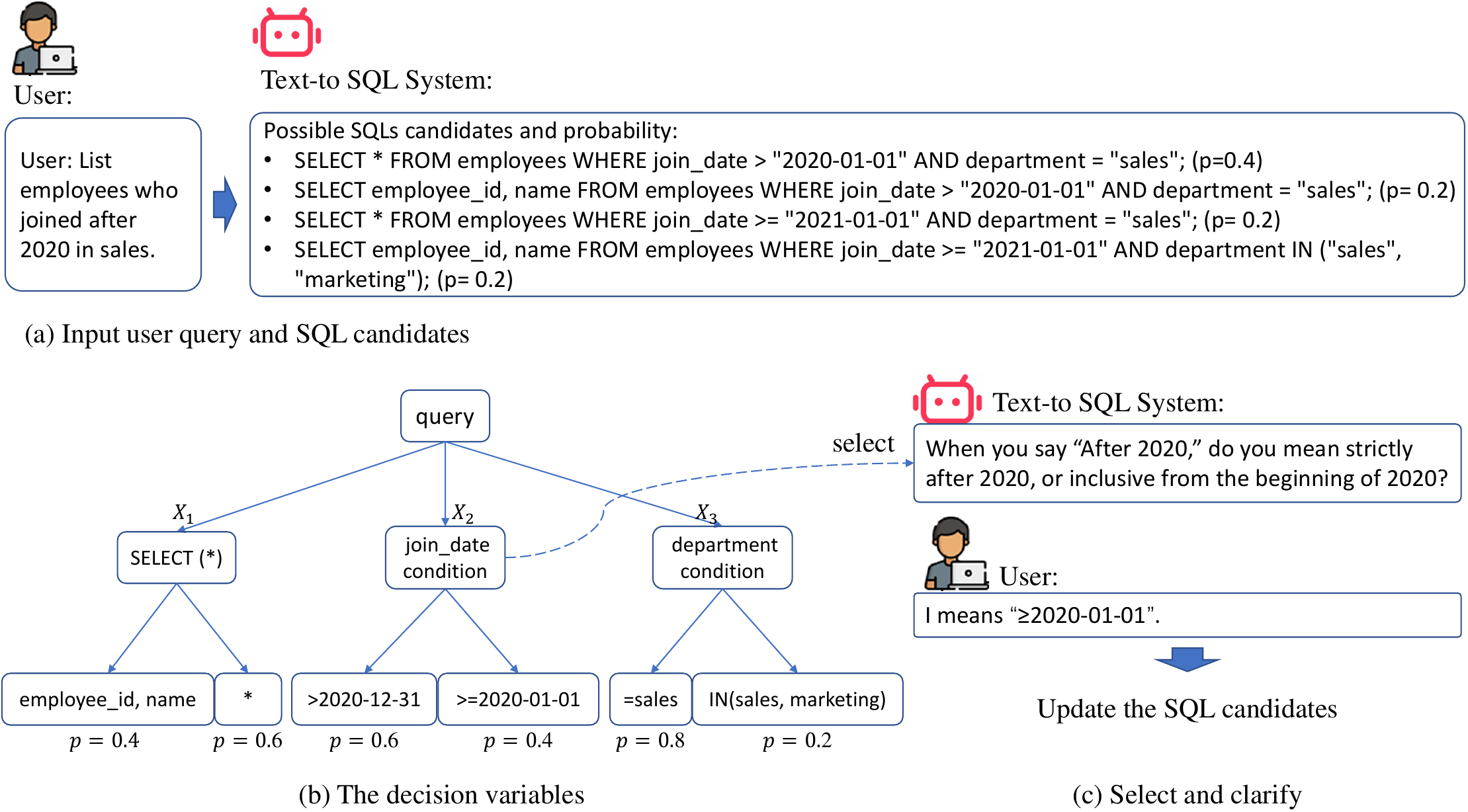}\\
\caption{Illustration of the (a) input user query and several candidate SQLs with possibility; and (b) the decision branches with probability; (c) the expected information gain.}
\label{fig:decision}
\end{figure*}

In practice, the interactive clarification mechanism leveraging EIG identifies the most impactful decision points, efficiently resolves ambiguities, and progressively refines the query to achieve high accuracy. Fig.~\ref{fig:decision} demonstrates this iterative refinement process with explicit calculations of information gain, guiding the interactive clarification systematically.

Thus, by formally modeling the standard and ambiguous scenarios explicitly, our formulation provides a rigorous theoretical foundation for interactive Text-to-SQL translation that significantly improves robustness and interpretability in handling real-world ambiguities.

\begin{figure}
\centering
\includegraphics[width=0.7\linewidth]{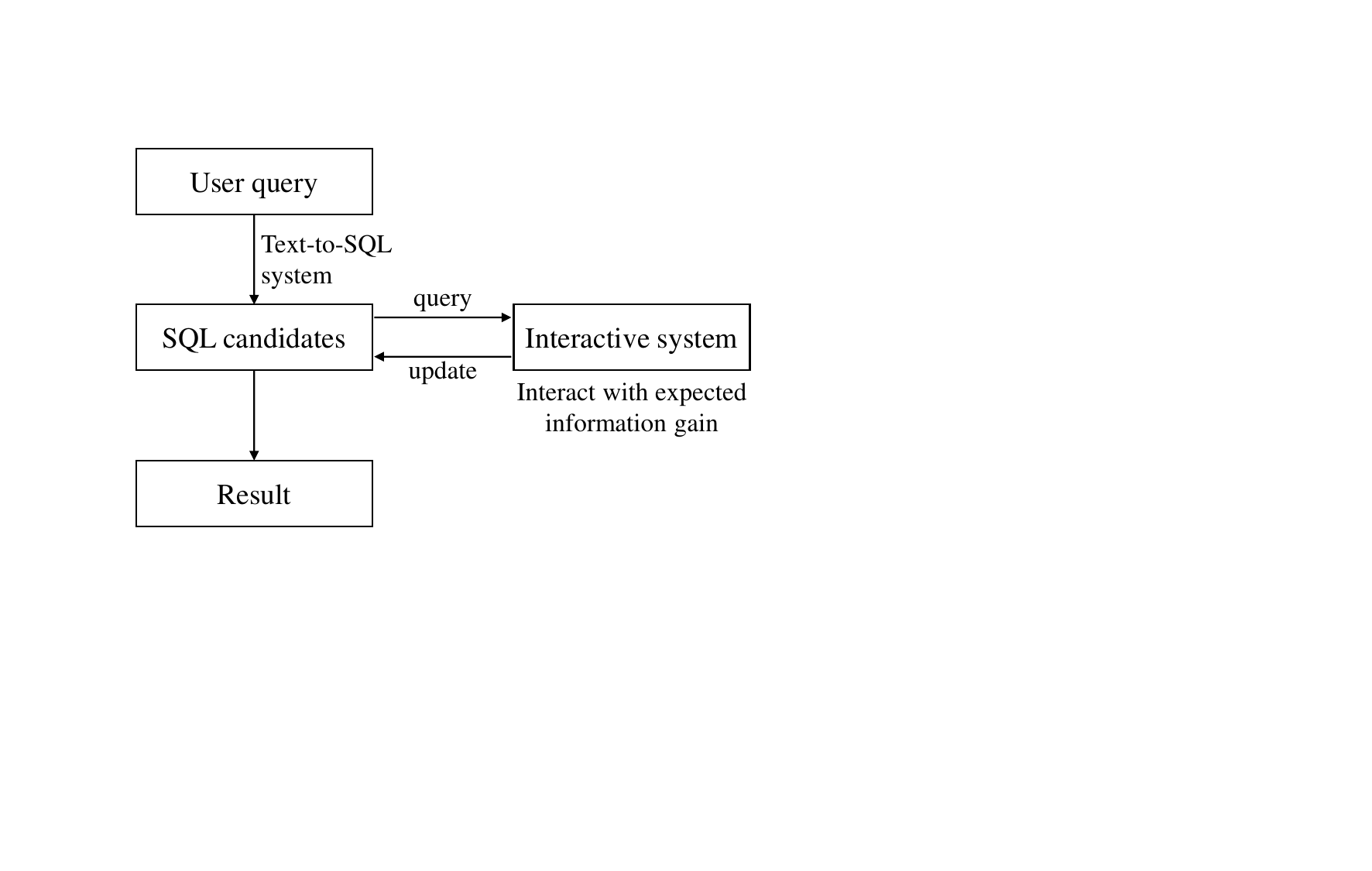}\\
\caption{The proposed interactive Text-to-SQL system.}
\label{fig:flow}
\end{figure}

\section{Solution}

In this section, we present a complete solution to select clarification to reduce the ambiguity of the input user query.

The overall pipeline of the proposed interactive Text-to-SQL system is illustrated in Fig.~\ref{fig:flow}.
First, we use the expected information gain as the metric to evaluate the ambiguous decision variables $\mathcal{X}$, and derive necessary formulas to enable the computation. 
Second, we select the decision variables with max expected information gain and clarify them. Third, we update the SQL candidates based on the clarification. This process repeats iteratively until all the ambiguous variables are sufficiently clarified, and the system can confidently return the intended SQL.

\subsection{Branch Selection via EIG}

In order to design an effective strategy for selecting the optimal decision variable, it is essential to define a metric to estimate the importance of decision variables before they are clarified. Since the final objective is to reduce ambiguity, we use the Expected Information Gain (EIG) as the metric.

For a set of SQL candidates $\mathcal{Y}$ and decision variable $\mathcal{X}$, the Expected Information Gain is defined as the expected reduction in uncertainty. 
For each decision variable $X_i$, the expected information gain can be computed according to Eq.~\ref{eq:gain}. This criterion selects the variable whose clarification would, in expectation, most reduce the system's uncertainty about the correct SQL interpretation.
The decision variable with max expected information gain is selected as:

\begin{equation}
    X^* = \operatorname{arg max}_{X_i} I(X_i, \mathcal{Y}),
\end{equation}

We illustrate an example of the decision-making process for the user query : "List employees who joined after 2020 in sales." in Fig.~\ref{fig:decision}.
As shown in the figure, there are four possible query candidates and corresponding probabilities.
\begin{itemize}
    \item[1] SELECT * FROM employees WHERE join\_date $>$ "2020-01-01" AND department = "sales" (p=0.4)
    \item[2] SELECT employee\_id, name FROM employees WHERE join\_date $>$ "2020-01-01" AND department = "sales" (p=0.2)
    \item[3] SELECT * FROM employees WHERE join\_date $\geq$ "2021-01-01" AND department = "sales" (p=0.2)
    \item[4] SELECT employee\_id, name FROM employees WHERE join\_date $\geq$ "2021-01-01" AND department IN ("sales", "marketing") (p=0.2)
\end{itemize}
In each level, there are multiple decision branches. We take Branch $X_3$ as an example to compute the information gain as:

\begin{equation}
    \begin{aligned}
        H(\mathcal{Y}) &= 1.922,\\
        H(\mathcal{Y}|X_3) &= 0.8 \times 1.5 + 0.2 \times 0 = 1.2,\\
        I(X_3;\mathcal{Y}) &= 1.922 - 1.2 = 0.722.
    \end{aligned}
\end{equation}
where $H(\mathcal{Y})$ is the initial entropy before branch $X_3$, $H(\mathcal{\mathcal{Y}}|X_3)$ is the conditional entropy, and $I(X_3;\mathcal{Y})=0.722$ is the mutual information. Similarly, we can compute the mutual information for branches $X_1,X_2$ and $X_4$, and select the branch which has max mutual information to clarify.


\subsection{Interaction and Query Refinement in Multi-turn Conversation}

Once the system selects a decision variable $X_i$, it formulates a natural language clarification question corresponding to the ambiguity. For example:
\begin{itemize}
    \item For a time ambiguity: “By ‘after 2020’, do you mean $>2020$ or $\geq 2020$?”
    \item For a vague noun phrase: “When you say ‘in sale’, are you referring to the department or the function?"
    \item For a schema ambiguity: “When you say ‘List employees’, do you mean list all column or only employee ID and name"
\end{itemize}
The user's response is mapped back to a concrete value $x^*$. The system then filters the candidate SQLs $\mathcal{Y}$, retaining only those consistent with the choice $X_i=x^*$, and renormalizes the probability $P(\mathcal{Y})$.
This process is repeated iteratively until:

\begin{itemize}
    \item The maximum candidate confidence exceeds a predefined threshold $\tau$, or
    \item All decision variables are resolved, and the distribution $P(\mathcal{Y})$ is unimodal.
\end{itemize}
The final SQL output $Q^*$ is returned as the model output:
\begin{equation}
    Q^* = \operatorname{arg max}_{Q\in \mathcal{Y}}P(Q).
\end{equation}
Fig.\ref{fig:flow} illustrates the pipeline of the system. In our implementation, we adopt a multi-turn interaction framework to better reflect the behavior of real-world database question-answering systems. 
For each input query, we generate multi-turn interaction data $T_{instruct}$ as,
\begin{equation}
T_{instruct} = (q, T^1_q, T^1_a, T^2_q, T^2_a,...),
\end{equation}
where $q$ is the input query by the user, and $T^i_q$ and $T^i_a$ denote the clarification question and answer by the user at $i$-th round, respectively.
This design allows the system to actively engage in a clarification conversation when ambiguity is detected and iteratively refine its understanding of the user’s intent.




\subsection{Finding the Optimal Query}

In the interactive Text-to-SQL framework proposed in this paper, each clarification step aims to reduce ambiguity by selectively querying the user about the interpretation of decision variables. As discussed previously, each interaction seeks to identify the decision point whose clarification maximizes the expected reduction in the uncertainty of the SQL candidate distribution \(\mathcal{Y}\). We formalize this selection criterion by leveraging the concept of expected information gain (EIG), measured by mutual information:

\begin{equation}
    I(X_i; \mathcal{Y}) = H(\mathcal{Y}) - H(\mathcal{Y} \mid X_i),
\end{equation}

where the entropy \(H(\mathcal{Y}) = -\sum_{y \in \mathcal{Y}} P(y) \log P(y)\) quantifies the uncertainty associated with the current candidate SQL distribution, and \(H(\mathcal{Y}\mid X_i)\) represents the conditional entropy of \(\mathcal{Y}\) after resolving ambiguity about decision variable \(X_i\).

As shown in our preceding sections, directly computing this EIG criterion for all candidate decision variables may incur significant computational overhead (\(O(NM)\)), where \(N = |\mathcal{Y}|\) is the number of SQL candidates and \(M = |\mathcal{X}|\) the number of decision variables. This complexity could hinder real-time responsiveness essential in practical systems. To address this challenge and align with our aim of deploying an efficient interactive system, we now present a theoretical result that significantly simplifies and accelerates the computation of optimal clarification points.

We first recall the branching structure concept introduced earlier, essential for understanding our approach.

\begin{definition}[Branching Tree]
Given a set of candidate SQL queries \(\mathcal{Y} = \{Q_1, Q_2, \dots, Q_N\}\) each associated with probabilities \(P(Q_i)\), and a set of decision variables \(\mathcal{X} = \{X_1, X_2, \dots, X_M\}\), the query disambiguation process can be naturally represented as a branching tree. Here, internal nodes correspond to decision variables \(X_i\), edges represent specific interpretations \(x \in \mathcal{V}_i\), and leaf nodes correspond to individual SQL query candidates. Thus, every path from the root to a leaf node uniquely specifies one candidate query interpretation.
\end{definition}

Exploiting this structured view of the interaction process, we derive the following theorem, facilitating efficient and optimal decision selection:

\begin{theorem}
Let \( Q^* = \arg\max_{Q \in \mathcal{Y}} P(Q) \) denote the candidate SQL query with the highest probability, and let \( \mathcal{P}(Q^*) = \{X_1, X_2, \dots, X_k\} \subseteq \mathcal{X} \) denote the sequence of decision variables along the path leading to \( Q^* \). Then, the optimal decision variable \( X^* \) is efficiently identified by the criterion:
\begin{equation}
    X^* = \arg\max_{X_i \in \mathcal{P}(Q^*)} P(Q^* \rightarrow X_i) \cdot H(X_i),
\end{equation}
where \( P(Q^* \rightarrow X_i) \) denotes the cumulative probability of candidate queries traversing through the decision node \( X_i \), and \( H(X_i) \) measures the entropy across its interpretations.
\end{theorem}

\begin{figure*}
\centering
\includegraphics[width=1\linewidth]{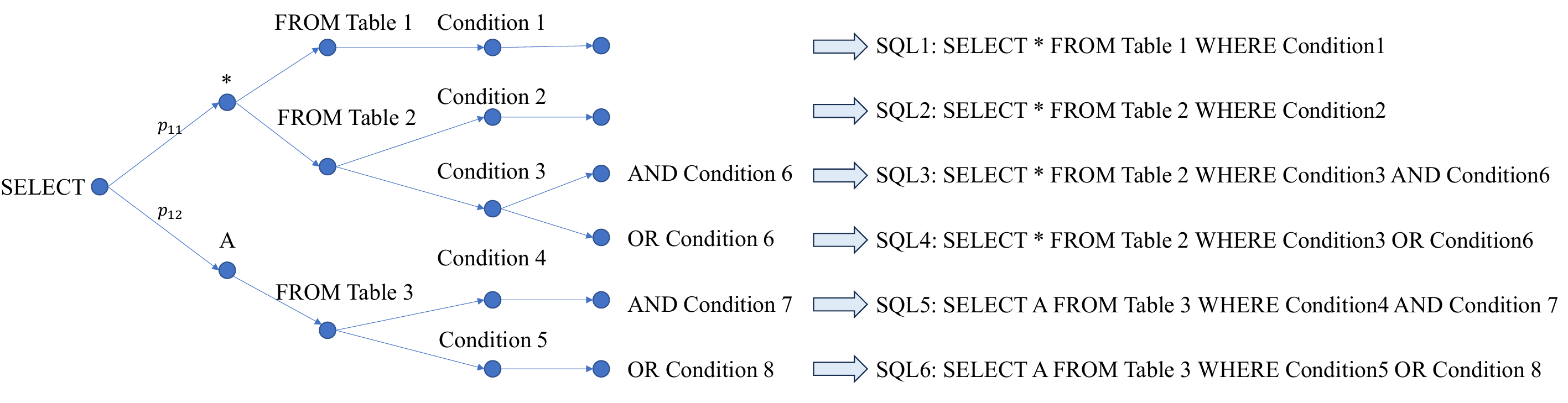}\\
\caption{Illustration of the graph representation of branch decisions process.}
\label{fig:token}
\end{figure*}

\begin{proof}
To rigorously prove this result, we begin with the definition of mutual information, explicitly expanded as:

\begin{align}
    I(X_i;\mathcal{Y}) &= H(\mathcal{Y}) - H(\mathcal{Y}\mid X_i) \nonumber \\
    &= H(\mathcal{Y}) - \sum_{x \in \mathcal{V}_i} P(X_i = x) H(\mathcal{Y}\mid X_i = x).
\end{align}

Note that the entropy reduction when clarifying a decision variable \(X_i\) is only influenced by candidate queries that pass through it, since those queries unaffected by the clarification do not alter the conditional distribution. Specifically, consider the decision variables lying on the path to the most probable candidate \(Q^*\). These variables partition the query distribution, isolating the subset of queries (including \(Q^*\)) that carry significant probability mass:

\[
P(Q^* \rightarrow X_i) = \sum_{Q \in \mathcal{Y},\, Q \rightarrow X_i} P(Q).
\]

Moreover, clarifying the decision variable \(X_i\) directly resolves the uncertainty regarding its interpretations. Hence, the conditional entropy at decision point \(X_i\) is precisely:

\[
H(X_i) = -\sum_{x \in \mathcal{V}_i} P(X_i = x)\log P(X_i = x).
\]

Combining these observations, the expected information gain simplifies elegantly to:

\[
I(X_i;\mathcal{Y}) = P(Q^* \rightarrow X_i) \cdot H(X_i).
\]

This completes the proof.
\end{proof}
Intuitively, this equation expresses that the optimal clarification choice balances two crucial factors:

\begin{itemize}
    \item \textbf{Relevance} (\(P(Q^* \rightarrow X_i)\)): the total probability mass of candidates influenced by clarifying \(X_i\).
    \item \textbf{Ambiguity} (\(H(X_i)\)): the degree of uncertainty regarding the interpretations at this decision point.
\end{itemize}

This trade-off ensures that the selected decision point is not only ambiguous (high uncertainty) but also influential (highly relevant to query selection).

From a computational perspective, the simplified form allows us to compute the optimal decision efficiently. Only one pass through the set of candidate SQL queries is necessary to calculate both the probability mass and interpretation entropy at each node along the path to \(Q^*\). Consequently, the computational complexity is reduced from \(O(NM)\) to \(O(N)\), significantly enhancing real-time responsiveness of our interactive approach.

In summary, this result fundamentally enhances our interactive framework, enabling computationally efficient, real-time clarification that systematically maximizes uncertainty reduction. We empirically validate the practical efficiency and effectiveness of this optimized decision selection criterion through extensive experimentation, as detailed in the following section.

\subsection{Implement Details}

In this section, we describe the implementation details of our proposed interactive disambiguation framework, including how SQL candidates are generated, how decision variables are extracted, and how clarification question are selected based on expected information gain. The detailed process is summarized in Algorithm 1, which will be introduced in the following parts.

\subsubsection{SQL Candidate Generation}
For each natural language query, we first generate a set of plausible SQL candidates using LLM, \eg, GPT-4. For each input natural language query, we ask the model to generate $n$ SQL queries $\mathcal{Y}\in \{Q_1,Q_2,...,Q_n\}$ along with their normalized generation probabilities $P(Q_i)$, where $Q_i\in \mathcal{Y}$. These probabilities reflect the model’s confidence over alternative interpretations and serve as the foundation for subsequent disambiguation.

\subsubsection{Branch Identification via Token-Level Comparison}
Given the set of SQL candidates $\mathcal{Y}$, we identify decision variables, \ie, semantic components of the query that vary across candidates and therefore represent potential ambiguities.
We tokenize each SQL query $\mathcal{Q}_i$ using a structural SQL-aware tokenizer (\eg, clause-level or AST-style) $\mathcal{Q}_i=\{t_i^1, t_i^2, ...,t_i^m\}$.
And then compare token across candidates to detect divergent elements. These typically include:
1) selected columns or aggregation functions, 2) WHERE clause conditions; 3) JOIN paths or intermediate tables, 4) GROUP BY or ORDER BY modifiers.

Each such variation corresponds to a decision variable $X_i$, with the set of observed alternatives forming its domain . For example, if some candidates filter by $\operatorname{start\_date} > "2021"$ and others use $\operatorname{start\_date} >= "2021"$, we create a decision variable for the filtering column with two possible values.

\subsubsection{Expected Information Gain Computation}
Once the set of branch variables $\mathcal{X}$ is extracted, the system selects one variable to clarify at each step. The selection is based on expected information gain (EIG), which quantifies how much uncertainty about the correct SQL query would be reduced by resolving each ambiguity.

For each decision variable $\mathcal{X}_i$, we estimate:
\begin{itemize}
    \item the marginal probability of each value $x\in V_i$, computed by summing the probabilities of SQL candidates that contain that value;
    \item the conditional distribution over SQL candidates given each possible value;
    \item the current entropy $H(Y)$ of the SQL candidate distribution, and the expected conditional entropy $H(Y|X_i)$.
\end{itemize}

The expected information gain for $X_i$ is then calculated as:
\begin{equation}
    I(X_i;\mathcal{Y}) = H(Y) - \sum_{x\in \mathcal{V}_i}P(x)H(Y|X_i=x)
\end{equation}
We select the decision variable $X^*=\operatorname{arg max}_{X_i} I(X_i;\mathcal{Y})$ for clarification. A natural language question is then generated to present the alternatives in $\mathcal{V}_i$ to the user.

\subsubsection{Post-Clarification Update}

After receiving the user’s response (\eg, selecting a preferred interpretation), the system filters out all SQL candidates inconsistent with that value and renormalizes the distribution over the remaining SQLs. This updated distribution becomes the basis for the next round of interaction.

The process repeats until either:
\begin{itemize}
    \item a candidate exceeds a confidence threshold, or
    \item no ambiguities remain in the candidate set.
\end{itemize}
Our implementation is modular and model-agnostic; it hence can be used on top of any SQL generation model that outputs ranked candidates and requires no retraining of the base model.

An example of the branch decision is illustrated in Fig.~\ref{fig:token}. As shown in the figure, the model generates six SQL candidates with associated probabilities. Based on the distribution over SQLs, the expected information gain is computed for each decision variable. 
From these, the system identifies several decision points, including,
\begin{itemize}
    \item Column choice: $*, A, B$;
    \item Table name: Table 1, ..., Table 4;
    \item Conditions: Condition 1,...,Condition 5.
\end{itemize}

Based on the distribution over SQLs, the expected information gain is computed for each decision variable. Suppose the column choice yields the highest EIG; the system then generates a question such as:

\begin{equation}
"Should\ the\ output\ group\ by\ B\ or\ not?"
\end{equation}
Once the user selects $B$, the system filters out SQLs using other columns, updates the distribution, and repeats this process until one candidate becomes dominant.

\section{Experiments}

\subsection{Datasets}

We evaluate our approach on two benchmark datasets, including Spider and AmbiQT.

\textbf{Spider:} Spider dataset~\cite{yu2018spider} is a widely-used cross-domain Text-to-SQL benchmark designed to test generalization to unseen databases. It contains 10,181 natural language questions annotated with gold SQL queries over 200 databases spanning 138 different schemas. Queries in Spider involve complex SQL operations, including joins, group-by, nested subqueries, and aggregations.

We use the standard Spider train/dev split and evaluate using exact match accuracy and execution accuracy, following common practice. While Spider was not designed specifically for ambiguity, we use it to validate our model’s performance in general Text-to-SQL translation and to measure whether interactive disambiguation incurs unnecessary overhead when queries are clear.

\textbf{AmbiQT:} AmbiQT~\cite{bhaskar2023benchmarking} is a recently introduced benchmark focusing on semantic ambiguity in Text-to-SQL. Each natural language query in AmbiQT is deliberately underspecified or contextually ambiguous, and is annotated with multiple correct SQL interpretations. The goal is to evaluate a model's ability to detect ambiguity and disambiguate through clarification.

The AmbiQT dataset consists of four types of ambiguity, including Column Ambiguity (C), Table Ambiguity (T), Join Ambiguity (J), and Precomputed Aggregates (P). Each entry is designed so that both alternatives have a similar relevance to the question, and a well-calibrated decoding method is expected to rank them close by in their outputs. 
This dataset is particularly suited for testing our interactive framework, which is explicitly designed to resolve ambiguity through user interaction. We simulate user responses using the gold disambiguation annotations provided by AmbiQT.

\textbf{Spider 2.0:} Spider 2.0 is an enhanced version of the widely used Spider benchmark for Text-to-SQL generation. It retains the cross-domain nature of the original Spider dataset~\cite{yu2018spider}, where each natural language query is paired with a complex SQL statement over a different database schema.
Compared to Spider 1.0, Spider 2.0 introduces several improvements to better reflect real-world challenges. These include refined question annotations, more consistent SQL labeling, and a broader coverage of SQL features such as nested queries, set operations, and explicit handling of nulls and default values. The dataset also includes additional examples with increased lexical and structural diversity, making it more robust for evaluating generalization and ambiguity resolution in semantic parsing.

\subsection{Evaluation Metrics}

We evaluate the performance of our method using the official metrics for each dataset: exact set match accuracy and execution accuracy.

Exact set match accuracy treats each clause as a set and compares the predicted SQL query to the reference query on a clause-by-clause basis. A prediction is considered correct only if all components exactly match the ground truth, excluding value comparisons.
Execution accuracy, in contrast, measures whether the predicted query yields the same execution result as the ground truth query on the database. This metric provides a more reliable assessment of model performance, as multiple SQL queries can be semantically equivalent even if they differ syntactically.

\begin{algorithm*}[t]
\caption{Interactive Text-to-SQL with Expeceted Information Gain Algorithm}
\KwIn{
Natural language query $q$,\\
\ \ \ \ \ \ \ \ \ \ Database schema $S$,\\
\ \ \ \ \ \ \ \ \ \ SQL candidates $\mathcal{Y} = \{Q_1, ..., Q_k\}$,\\
\ \ \ \ \ \ \ \ \ \ probability for each SQL candidate $P(Q_i)$.}
\KwOut{Final SQL query $y^*$}

\While{True}
{
Detect a set of decision variables $\mathcal{X}=\{X_1,X_2,...,X_M\}$ to clarify based on token-level differences in $\mathcal{Y}$.

\If{$\mathcal{X}$ is $\operatorname{None}$}{break}

Compute information entropy at decision point $X$.\\

\For{$i\leftarrow 1$ \KwTo $M$}
{
Compute information entropy for $X_i$.\\
Compute the expected uncertainty after clarifying branch $X_i$: $ H(Y|X_i)$.\\
Compute the expected information gain $I(X_i; Y) = H(Y) - H(Y | X_i)$
}
Select $x^* = \arg\max \text{EIG}(X_i)$\;
Ask clarification question for $x$ and obtain user response $x^*$\;
Filter candidates: $\mathcal{Y} \leftarrow \{Q_i \in \mathcal{Y} : y_i[X] = x^*\}$\;
Renormalize $P(Q_i)$ over updated $\mathcal{Y}$\;
Update decision variables $\mathcal{X}$\;

}
\Return{$y^* = \arg\max_{y_i \in \mathcal{Y}} P(y_i)$}
\label{alg:alg}
\end{algorithm*}

\subsection{Ablation Study}

\subsubsection{Evaluation of EIG Selection:}

In this section, we verify the effectiveness of our proposed method by comparing it with several alternative strategies for selecting clarification questions during the interaction process on the Spider dataset. All variants are evaluated under the same framework, where a set of SQL candidates is maintained and interaction is performed to resolve semantic ambiguities.

We compare the following five strategies:

\begin{itemize}
    \item Random Selection: At each step, we randomly select a decision variable for clarification. This approach serves as a naive interactive baseline without reasoning about informativeness.
    \item Max Probability First: The decision variable with max probablity across SQL candidates is selected first. This heuristic assumes that resolving dominant patterns may lead to best result.
    \item Min Probability First: This variant prioritizes decision variables with rare or diverse values across candidates, under the assumption that resolving disagreement early may eliminate more uncertainty.
    \item Information Gain (IG): Selects the decision variable with the largest raw information gain (i.e., difference in entropy before and after clarification), assuming a uniform prior over values. This strategy measures informativeness but ignores the actual distribution over user intent.
    \item Expected Information Gain (EIG): Our method selects the decision variable with the highest expected information gain, computed under the current distribution over SQL candidates. It considers both the informativeness and the likelihood of different outcomes, resulting in globally optimal decision-making.
\end{itemize}

Considering that most samples in the Spider dataset contain relatively little ambiguity, we construct an ambiguous subset of Spider to better evaluate our method’s ability to resolve uncertainty. Specifically, we use the model's confidence scores over multiple SQL candidates to estimate the degree of ambiguity in each input query.
For each example, we generate a set of candidate SQL queries using beam search, along with their normalized probabilities. If the highest-probability candidate has a confidence score lower than 0.7, we regard the input query as ambiguous and include it in our evaluation subset. The underlying intuition is that when the model fails to strongly prefer one query over others, it reflects genuine semantic uncertainty in the input, such as underspecified conditions, vague references, or multiple plausible interpretations.

\begin{table*}
\begin{center}
\caption{Exact match and Execution accuracy on Spider dataset.}
\setlength{\tabcolsep}{20 pt}
\small
\begin{tabular}{l|cccc|c}
\toprule
Type &Easy &Medium &Hard &Extra  &All     \\
\midrule
Num       &26 &91 &87 &86 &290 \\
\midrule
\multicolumn{6}{c}{\emph{Exact match}}\\
\midrule
Random Select    &30.8 &33.0&10.6&7.0 &19.0\\
Min Probability  &23.1 &14.3&8.0 &2.3 &9.7 \\
Max Probability  &50.0 &47.1&30.3&14.0&30.6\\
Info Gain        &50.0 &49.5&33.3&12.8&33.8\\
\midrule
Expected Info Gain  &\textbf{57.7}&\textbf{53.8}&\textbf{40.2}&\textbf{14.0}&\textbf{38.3}\\
\midrule
\multicolumn{6}{c}{\emph{Execution}}\\
\midrule
Random Select    &64.1 &79.9 &68.2 &54.6 &70.1 \\
Min Probability  &61.5 &79.1 &66.7 &51.2 &65.5 \\
Max Probability  &65.4 &82.4 &74.7 &60.5 &72.1 \\
Info Gain &73.1 &83.5 &73.6 &62.8 &73.4 \\
\midrule
Expected Info Gain  &\textbf{76.9} &\textbf{86.8} &\textbf{77.0} &\textbf{65.1} &\textbf{76.6} \\
\bottomrule
\end{tabular}
\label{table:spider}
\end{center}
\end{table*}

This filtered subset allows us to test the effectiveness of our interactive disambiguation framework under more realistic and challenging conditions, even within a benchmark like Spider that was not explicitly designed to model ambiguity.
Tables~\ref{table:spider} reports the results of various interaction strategies on this subset in terms of execution accuracy and exact match accuracy, broken down by question hardness levels (Easy, Medium, Hard, Extra Hard).
Across both metrics, our proposed method significantly outperforms all baselines. In terms of execution accuracy, our method achieves 76.6\% overall, a clear improvement over the best baseline (Info Gain, 73.4\%) and heuristic methods such as Max Probability (72.1\%) and Random (70.1\%). The gains are especially notable on harder examples, with 77.0\% on Hard and 65.1\% on Extra Hard questions, compared to 73.6\% and 62.8\% for Info Gain, respectively.

For exact match accuracy, which reflects the model's ability to produce structurally and semantically precise SQL queries, the improvement is even more substantial. Our method achieves 38.3\% overall, compared to 21.0\% for Info Gain and only 19.0\% for Max Probability. The performance gap widens further on challenging cases: for Hard questions, we reach 40.2\%, more than double that of Random (6.6\%) and Min Prob (8.0\%). On Easy and Medium questions, we observe strong gains as well, with 57.5\% and 53.8\%, respectively.
These results validate that our expected information gain-based strategy not only helps disambiguate user intent more accurately, but also translates into concrete improvements in both execution correctness and exact query structure. Compared to baseline heuristics, which either prioritize frequency or ignore probabilistic impact, our method consistently asks more relevant questions and converges toward the intended query with fewer, more meaningful interactions.

The performance gap is particularly prominent on non-trivial queries, where ambiguity tends to be higher and the space of possible interpretations more complex. This suggests that our method is especially effective in resolving deep semantic uncertainty, making it a strong candidate for real-world deployment where users often submit vague or conversational queries.

\begin{table}
\begin{center}
\caption{Evaluation of the multi-turn conversation on Spider and AmbiQT dataset.}
\setlength{\tabcolsep}{5.5 pt}
\small
\begin{tabular}{l|cc|cc}
\toprule
\multirow{2}{*}{Method} &\multicolumn{2}{c|}{Spider} &\multicolumn{2}{c}{AmbiQT}     \\
\cmidrule{2-5}
       & Exact match &Execution &Exact match &Execution  \\
\midrule
Single-turn &36.7 &75.1 &68.13 &63.92\\
Multi-turn  &38.3 &76.6 &69.72 &65.26 \\
\bottomrule
\end{tabular}
\label{table:multiturn}
\end{center}
\end{table}

\subsubsection{Evaluation of Multi-turn Conversation:}

To better understand the effectiveness of interactive disambiguation, we compare two variants of our framework: single-turn interaction and multi-turn interaction.
In the single-turn setting, the system asks one clarification question at a time but does not retain the history of prior questions and answers. Each question is selected based only on the current SQL distribution, without conditioning on previous interactions.
In the multi-turn setting, the system maintains a dialogue history, allowing it to take past clarifications into account when selecting the next ambiguity to resolve. This mirrors real-world conversational systems that adapt based on prior user input.

As shown in Table~\ref{table:multiturn}, multi-turn interaction significantly outperforms single-turn across both execution accuracy and exact match metrics. This demonstrates that tracking interaction history is crucial for resolving layered or dependent ambiguities in natural language queries. Without context, single-turn systems may ask redundant or suboptimal questions, limiting their ability to efficiently converge on the correct interpretation.

\begin{table*}
\caption{The top-$1$ exact match and execution accuracy on AmbiQT dataset.}
\setlength{\tabcolsep}{11 pt}
\small
\begin{tabular}{l|cccc}
\toprule
Type  &Column Ambiguity (C)  &Table Ambiguity (T)  &Join Ambiguity (J)  &Precomputed Aggregate (P)   \\
\midrule
\multicolumn{5}{c}{\emph{Exact match}}\\
\midrule
GPT          &35.32 &29.78&27.43&37.62\\
Codex        &35.24 &31.19&60.42&48.51\\
Flan-t5      &47.74 &45.94&83.68&58.42\\
T5-3B        &48.06 &41.99&78.82&68.32\\
LogicalBeam  &48.06 &42.06&78.82&68.32\\
\midrule
Ours   &\textbf{58.95} &\textbf{57.81} &\textbf{86.88} &\textbf{75.24}\\
\midrule
\multicolumn{5}{c}{\emph{Execution}}\\
\midrule
GPT          &43.47 &42.48&34.38&40.59\\
Codex        &39.03 &36.63&65.97&50.50\\
Flan-t5      &48.55 &46.65&82.29&41.58\\
T5-3B        &48.06 &44.39&76.04&50.50\\
LogicalBeam  &48.06 &43.47&76.04&50.50\\
\midrule
Ours  &\textbf{59.35} &\textbf{58.95}  &\textbf{83.33} &\textbf{59.41}\\
\bottomrule
\end{tabular}
\label{table:AmbiQT}
\end{table*}

\subsection{Comparison}

\subsubsection{Comparison on AmbiQT Dataset:}

We then compare the proposed Interactive EIG on the AmbiQT~\cite{bhaskar2023benchmarking} dataset. Different from the original test setting proposed in LogicalBeam~\cite{bhaskar2023benchmarking} which reports the top-5 accuracy on the AmbiQT dataset, we compare our method with recent models using a more challenging top-1 setting.
Table~\ref{table:AmbiQT} presents the exact match and execution accuracy for four major ambiguity categories defined in AmbiQT: Column Ambiguity (C), Table Ambiguity (T), Join Ambiguity (J), and Precomputed Aggregate (P). These categories represent different types of underspecification in natural language queries and allow for fine-grained evaluation of disambiguation capability.

Across all ambiguity types, our method consistently achieves the best performance. In terms of exact match accuracy, our model outperforms all compared methods on four ambiguity settings. Similar improvements are observed in execution accuracy, where our method also surpasses prior systems.
The superior performance of our approach highlights the importance of explicitly modeling ambiguity and engaging users to clarify intent. Notably, the improvements are especially significant in column and table ambiguity, where surface-level similarity and overlapping schema names make one-shot prediction unreliable. By identifying these ambiguity sources and resolving them interactively via targeted questions, our system can recover accurate interpretations that static models often miss.
In summary, the results on AmbiQT demonstrate that our interactive disambiguation approach delivers substantial improvements across diverse ambiguity categories, confirming its robustness and general applicability in real-world Text-to-SQL scenarios.

\subsubsection{Comparison on Spider2 Dataset:}

To assess the generalizability of our approach, we further evaluate its effectiveness on the Spider 2.0 dataset. Table~\ref{table:spider2} presents the execution accuracy of several baseline methods, both with and without our proposed Expected Information Gain (EIG)–based interaction mechanism.

Across all evaluated systems—DIN-SQL, DAIL-SQL, and Spider-Agent—the incorporation of EIG leads to consistent performance improvements. For instance, DIN-SQL sees a relative gain from 1.46\% to 4.41\%, while Spider-Agent improves from 13.71\% to 17.12\%. These results indicate that our method provides complementary benefits to existing Text-to-SQL models, regardless of their underlying architecture.
Even LinkAlign, a strong alignment-based baseline, achieves a significant improvement, increasing from 24.86\% to 27.91\% when augmented with our EIG strategy.
Overall, the improvements demonstrate that our EIG-based disambiguation strategy is effective even on more realistic and challenging queries from Spider 2.0, reinforcing the practical applicability of our method beyond synthetic or benchmark-specific settings.

\begin{table}
\begin{center}
\caption{Evaluation of the proposed method on Spider2 dataset.}
\setlength{\tabcolsep}{20 pt}
\small
\begin{tabular}{l|c|ccc}
\toprule
Method       &w/o EIG &w/ EIG  \\
\midrule
DIN-SQL  &1.46 &4.41\\
DAIL-SQL &5.68 &8.56\\
Spider-Agent &13.71 &17.12\\
LinkAlign    &24.86 &27.91\\
\bottomrule
\end{tabular}
\label{table:spider2}
\end{center}
\end{table}

\section{Related Work}

\subsection{Text-to-SQL Generation}
Text-to-SQL generation task aims to translate natural language questions into executable SQL queries over structured relational databases.
The construction of several large text-to-SQL datasets, such as WikiSQL~\cite{zhong2017seq2sql} and Spider~\cite{yu2018spider}, has enabled the adoption of deep learning models in this task, achieving unprecedented performance in recent years.

Early models adopted the sequence-to-sequence paradigm~\cite{sutskever2014sequence}, jointly encoding the question and database schema and decoding the SQL output. Various encoder architectures have been explored, including BiLSTMs in IRNet~\cite{kawakami2008supervised}, CNNs in RYANSQL~\cite{choi2021ryansql}, pretrained language models like BERT in SQLova~\cite{hwang2019comprehensive}, and graph neural networks in RATSQL~\cite{wang2019rat}. Other work introduces intermediate representations~\cite{gan2021natural} or uses table-aware pretraining, such as TaBERT~\cite{yin2020tabert}, TaPas~\cite{herzig2020tapas}, and Grappa~\cite{yu2020grappa}.
Decoder-side methods can be categorized into sketch-based and generation-based approaches~\cite{qin2022survey}. Sketch-based methods predict SQL slots using predefined templates~\cite{xu2017sqlnet,hwang2019comprehensive}, but often struggle with compositional generalization. Generation-based models~\cite{guo2019towards,wang2019rat,cao2021lgesql} output SQL queries as syntax trees, offering greater flexibility.

More recently, prompting-based methods have emerged, leveraging large language models (LLMs) for zero- or few-shot SQL generation. Rajkumar et al.\cite{rajkumar2022evaluating} and Liu et al.\cite{liu2023comprehensive} show that LLMs can perform well on the Spider benchmark using carefully designed prompts and schema serialization. These methods avoid task-specific training and have also been extended to related table reasoning tasks~\cite{guo2023few,chen2022large}. While powerful, LLMs remain sensitive to prompt formulation and struggle with schema grounding and ambiguous queries.

\subsection{Interactive Text-to-SQL Generation}

There is growing interest in interactive approaches that incorporate user feedback to improve SQL generation. Iyer et al.\cite{iyer2017learning} proposed allowing users to flag incorrect queries to retrain the model. DIY\cite{narechania2021diy} and NaLIR~\cite{li2014constructing} let users revise queries by selecting alternative values or expressions. Later systems such as PIIA~\cite{li2020you}, MISP~\cite{yao2019model}, and Dial-SQL~\cite{gur2018dialsql} proactively query users through multiple-choice clarification prompts.

Despite these advances, prior interactive methods often lack a principled strategy for resolving ambiguity. Many follow fixed decision orders (e.g., SELECT before WHERE) or rely on heuristic rules, which can result in inefficient interactions and leave core ambiguities unresolved.

In contrast, our approach maintains a distribution over plausible SQL interpretations and uses expected information gain to select the most informative clarification at each step. This enables globally optimized interaction, reducing user burden while ensuring rapid disambiguation and improved transparency.

\subsection{Ambiguity in Text-to-SQL Generation}

Ambiguity is a well-known challenge across natural language processing (NLP). It has been extensively studied in tasks such as summarization, vision-language grounding, and dialogue systems. Pilault et al.\cite{pilault2023interactive} propose an interactive summarization framework that detects and clarifies underspecified content through user engagement. VISa\cite{li2022visa} resolves referential ambiguity in multimodal settings, while Futeral et al.~\cite{futeral2022tackling} tackle unclear user intents in task-oriented dialogue via clarification questions. These works treat ambiguity as an essential modeling signal and highlight the value of interaction for resolving uncertainty when user intent is underspecified.

In contrast, ambiguity has received limited attention in the Text-to-SQL literature. Most existing methods assume queries are fully specified and directly map to a single correct SQL output. However, real-world questions are often vague, context-dependent, or underdetermined. Wang et al.~\cite{wang2022know} address column ambiguity by labeling token spans, but their method is limited in scope and does not generalize to other ambiguity types such as joins or temporal expressions.

To better evaluate systems under ambiguity, Bhaskar et al.~\cite{bhaskar2023benchmarking} introduce AmbiQT, a benchmark containing diverse ambiguous queries and multiple valid SQL interpretations, along with corresponding clarification questions. While AmbiQT exposes key challenges, most existing approaches evaluated on it rely on static decoding or reranking, without principled interaction.

In this work, we propose a general and interactive disambiguation framework that explicitly models uncertainty over the SQL space. By computing the expected information gain of candidate clarification points, our method selects the most informative questions and engages the user to resolve ambiguity efficiently. This strategy enables broader generalization across ambiguity types and aligns with growing interest in interactive and user-centered NLP systems.

\section{Future work}

Despite the progress made in this study, there remain several promising directions for future exploration to improve its robustness and deployability in real-world scenarios.
(1) Future systems should generate clarification questions that better reflect the user’s original phrasing, improving fluency and understanding.
(2) Real users may provide incorrect or ambiguous responses. Simulating such behavior through noisy user models can help evaluate and improve system robustness.
(3) Moving beyond isolated queries, future work can incorporate user history, roles, and preferences to resolve ambiguity more effectively in real-world applications.
By addressing these directions, we aim to build Text-to-SQL systems that are not only accurate in benchmark settings but also resilient, adaptive, and intuitive enough for practical deployment.

\section{Conclusion}

This paper presents an interactive Text-to-SQL framework that addresses semantic ambiguity in natural language queries by modeling SQL generation as probabilistic reasoning over multiple candidate queries. Instead of assuming fully specified input, the system maintains a distribution over plausible interpretations and resolves uncertainty through targeted user interaction. At each step, it selects the most informative clarification based on expected information gain, enabling efficient disambiguation with minimal user effort.
Experiments on the AmbiQT benchmark and an ambiguous subset of Spider demonstrate that our method consistently outperforms baselines in both execution and exact match accuracy, particularly on complex and ambiguous queries.
These results highlight the value of principled, interaction-driven disambiguation in Text-to-SQL tasks. Our framework is simple, adaptable, and model-agnostic, offering a promising direction for building more accurate and user-aligned database interfaces. Future work includes extending to multi-turn clarification and learning-based question generation.


\bibliographystyle{ACM-Reference-Format}
\bibliography{sample}

\end{document}